\documentstyle[aps,psfig]{revtex}
\begin{document}

\tightenlines
\draft

\pagestyle{plain}
\sloppy
\baselineskip 0.5cm

\title{Phase-Disorder Effects in a Cellular Automaton Model of Epidemic
Propagation}

\author{M. Bezzi $^{1}$ and  R. Livi$^{2}$}  
\vspace{1.cm}

\address{$(1)$ Dipartimento di Fisica dell' Universit\`a and 
Sezione I.N.F.N. di Bologna, Via Irnerio 46, 40126 Bologna, Italy\\ 
$(2)$ Dipartimento di Fisica dell' Universit\`a , Sezione I.N.F.N. and 
   Unit\`a  I.N.F.M. di Firenze, L.go E. Fermi 2, 50125 Firenze, Italy\\   
}   
\date{\today}
\maketitle

\begin{abstract}
A deterministic cellular automaton rule defined on the Moore neighbourhood 
is studied as a model of epidemic propagation.  The directed nature of the
interaction between cells allows one to introduce
the dependence on a disorder 
parameter that determines the fraction of ``in-phase'' cells.  
Phase-disorder is shown to produce peculiar changes in the dynamical
and statistical properties of the different evolution regimes obtained
by varying the infection and the immunization periods.
In particular, the finite-velocity spreading of perturbations,
characterizing chaotic evolution, can be 
prevented by localization effects induced by phase-disorder, that 
may also yield spatial isotropy of the infection propagation as
a statistical effect. Analogously, the structure of phase-synchronous 
ordered patterns is rapidly lost as soon as phase-disorder is increased,
yielding a defect-mediated turbulent regime . 
\end{abstract}
\vskip 1.cm

\section{Introduction}
Deterministic Cellular Automata (DCA) have revealed very 
useful tools for characterizing some basic  
complex behaviours emerging from simple, local evolution rules.
On the other hand, they are not particularly suitable for
applications, mainly because they are independent of   
some tunable parameter. This is why the mechanisms of 
front propagation in epidemics, forest fires
and chemical oscillations are usually modelled on the basis of probabilistic
CA rules (see, for instance \cite{Schon,Bocc}). 
Probability plays the role of a control parameter,
that may yield interesting phenomena like phase transitions between
different dynamical regimes (e.g., spiral-wave versus fully developed turbulence
or extinction versus ordered front propagation).

A similar and equally interesting scenario has been recently
obtained for a 3-state DCA rule of directed
epidemic propagation in 2d~\cite{Rous}~.
In this model the state-space of each cell is extended to a 
phase variable that allows for introducing  
a ``quenched-disorder'' parameter.
 
The complete definition of the model introduced in \cite{Rous} 
and a
summary of the main results therein obtained is contained 
in Section 2.  The extension of the model to nearest- and 
next-to-nearest neighbour interaction 
is discussed in Section 3, where  
the different dynamical regimes, characterizing
the ``phase-synchronous'' case, are also enlisted. The main difference
with respect to the model considered in\cite{Rous}, where the
interaction was limited to nearest-neighbour cells, 
is that the mechanism of infection propagation in the chaotic 
regime is characterized by two typical speeds, associated to the 
anysotropy of the interaction induced by the lattice.  
Section 4 is devoted
to a detailed account of the effects induced by quenched
phase-disorder on both chaotic and ordered dynamical regimes.
In particular, we show that
there are chaotic regimes where 
sufficiently small values of the phase-disorder paramter can inhibit the
spreading of perturbations at finite velocity. If this localization effect
is absent, one obtains that, at some finite value of the  
phase-disorder parameter, the isotropy of the infection mechanism 
can be restored as a statistical effect. 
Moreover, we find that the structure of phase-synchronous 
ordered patterns is rapidly lost as soon as $p$ is decreased.
Such a phenomenon is strongly reminiscent of the transition  
from a weakly-turbulent regime to a strongly-turbulent one,
already abserved in lattices of coupled stable maps \cite{Cuche}.
  
Conclusions are drawn in Section 5.
 
\section{ The Model} 
 
The DCA model is defined on a square lattice of size $L\times L$ 
with periodic boundary conditions. Each cell can assume three 
different states : infected ($I$), susceptible ($S$) and immunized, 
or quiescent ($Q$). Its neighbourhood is made of ${\cal N}$ cells.
The model depends on two main parameters:
$T_I$ and $T_Q$, the infection and immunization periods, respectively.
As we have already pointed out in the introduction, each cell
depends also on a phase variable $\theta = 1,\cdots,{\cal N}$.
At each instant of time $t$, any cell interacts with only one of the
cells in its neighbourhood, that is selected by the value of   
$\theta (t)$.   

The deterministic updating rule is defined as follows :
\begin{itemize}
\item[\bf R1]{If the phase variable of an $I$-cell selects a $S$-cell,
this becomes an $I$-cell at the following time step.} 
\item[\bf R2]{The infection (immunization) time of an $I$- ($Q$) -cell is
increased by a unit time step, so that after a period of $T_I$ ($T_Q$)
steps an $I$- ($Q$)-cell becomes a $Q$- ($S$)-cell.}
\item[\bf R3]{Independently of the state of the cell, its phase variable
is increased by one unit. Accordingly, as time flows $\theta$ spans 
sequentially all the cells in the neighbourhood, turning back
to its initial value after ${\cal N}$ time steps.}
\end{itemize}
The presence of the phase variable $\theta$ allows one to introduce
the quenched-disorder parameter $p$, measuring the fraction of cells 
sharing the same initial phase value $\theta(0)$.
According to {\bf R3}, these cells will then  share the same
phase value at each instant of time.
In the ``phase-synchronous'' case, $p \,=\, 1$, all cells have the 
same phase value. For $ 1/{\cal N}\leq p \leq 1$, a fraction of 
$p$ randomly chosen cells have the same phase value, while the remaining 
cells equally and randomly share the other ${\cal N} -1$ phase values.

This 2-d epidemic propagation model
was studied for ${\cal N} = 4 $ (Von Neumann neighbourhood) 
by Giorgini et al. \cite{Rous}. 
Let us summarize the main results, introducing the same symbolic 
notations used in the quoted reference.

First of all the authors classified the different dynamical regimes 
characterizing the phase-synchronous case ($p = 1$), when both $T_I$ 
and $T_Q$ are varied. 
To this aim, they considered a 
specific initial condition with a single $I$-cell in a ``sea'' of
$S$-cells.
For small values of  $T_I$ (smaller than or equal to ${\cal N}/2$)
and for $T_Q$ large enough (at least larger or equal to 
${\cal N}/2$), the epidemy, after a transient period of time, 
dies out (extinction regime, symbolically denoted by E). Actually, 
a cluster of $I$-cells 
develops quite soon, but its further propagation is inhibited by the
$Q$-cells that, after a sufficient time span, form at its boundary.
For smaller values of $T_Q$, the infection
remains confined inside a finite lattice region centered around 
the initially infected cell: the cluster of $I$-cells is 
characterized by a localized periodic evolution (symbol P). Note that
now the extinction of the infection does not occur, because
the periodicity of the phase variable allows for the re-infection of
previously infected cells, that had already turned to $Q$-cells,
according to {\bf R2}.
For large values of $T_I$ and $T_Q$,
the propagation of ordered fronts sets in (symbol F). 
The re-infection effect is absent and $I$-cells propagate
indefinitely as an ordered front, preserving the lattice 
square symmetry. It is straightforward to realize  
that in a finite size lattice such a front propagation dies out
after reaching the boundaries.

The interference of propagating fronts yields 
a seemingly chaotic evolution (symbol D), in which both the basic mechanism
(front propagation and re-contamination) are present and produce space time
patterns that do not exhibit any regularity.   
Note that such an outcome cannot be considered as totally unexpected, 
since it has been shown that, according to very general hypotheses, DCA 
rules with a state space larger than boolean may give rise to such 
non-trivial behaviours \cite{Grin}.

The quantity used in \cite{Rous} for obtaining a first characterization of 
the statistical properties associated to the D-regime was 
\begin{equation}
b(t)\,=\,{{\delta (t)}\over{I(t)\, S(t)}} 
\end{equation}
where $\delta (t)$ is the number (normalized to the lattice size
$L^2$) of $I$-cells interacting at time
$t$ with $S$-cells according to {\bf R1}, while $I(t)$ and $S(t)$
represent the ratios of infected and susceptible cells at the same 
time, respectively.
In practice, $b(t)$ can be 
interpreted as a measure of the rate of propagation of information 
through the lattice, or, equivalently, of the inhomogeinity of the
cell-state distribution. It was found to 
exhibit robust statistical properties in the D regime (self-averaging):
the hystogram of the frequency of occurrence of $b$ values vs. $b$ (i.e., the 
normalized probability distribution $P(b)$~) 
is a gaussian-like function.
 
It is worth stressing that any DCA rule must 
eventually yield a periodic evolution on a finite-size lattice. 
Nonetheless, one can conjecture that the ``transient'' chaotic
evolution typical of the D-regime would last forever in the 
thermodynamic limit ($L \rightarrow \infty$). Following \cite{PKLO},
this could be directly verified by measuring how the average duration of 
the D-regime increases with $L$. Since such a strategy is practically
impossible to be worked out satisfactorily for large enough 2-d samples,
the effective unpredictability of the D-regime can be detected indirectly
by measuring a finite average damage 
spreading velocity $\langle v_d \rangle$. A reliable 
determination of its value can be obtained by the following procedure.
One can start from a randomly seeded initial condition, obtained by assigning 
to all cells some normalized probabilities of being in the states $S$, 
$I$ and $Q$.  After a sufficiently long transient time,
any initial condition evolves to an irregular space-time pattern 
whose statistical features (e.g., $P(b)$) are found to be independent 
of the initial condition. 
A spatially localized perturbation of finite amplitude is introduced
in one of these patterns and the evolution of the difference field
with respect to the unpertubed pattern is then determined by measuring
at each instant of time $t$ the radius $ R(t) $ of the smallest circle 
where the difference field is non-zero. The averaging over a suitable 
set of initial conditions yield a reliable determination of 
\begin{equation} 
\langle v_d \rangle = {{d \langle R(t)\rangle}\over{dt}}
\label{voft}
\end{equation} 
For $p=1$ one observes $\langle v_d \rangle\not= 0$ in the D-regime;
decreasing $p$, $\langle v_d \rangle$ is found to approach zero 
at some finite value $p_c$, indicating the occurrence of a continuous
transition to a dynamical phase, where, as an effect of the phase-disorder,
perturbations remains localized.

\section{ The Moore-neighbourhood case}

We have extended the model described above to 
next-to-nearest neighbour interaction (${\cal N}=8$). 

Let us point out the main differences of this model 
with respect to the case studied in \cite{Rous}:
\begin{itemize}
\item[\bf a]- two velocities of propagation
of the infection mechanism along the main axes (nearest neighbour interaction) 
and along the diagonals (next-to-nearest neighbour interaction) of the square 
lattice are present;
\item[\bf b]- the commensurability relations of ${\cal N}$ 
with $T_I$ and $T_Q$ have changed. 
\end{itemize}

As we shall see, such features introduce some peculiar changes.
On the other hand, the basic mechanisms of re-contamination (infection
of a previously infected cell)
and front propagation
are found to play the crucial role also in this case. 

This is confirmed by the study of 
the dynamical regimes characterizing this DCA rule for $p = 1$,
when $T_I$ and $T_Q$ are varied.
Numerical simulations have been performed for $L = 400$, starting
from the initial condition of one $I$-cell in a ``sea'' of
$S$-cells \footnote{ For the sake of space, here we do not show 
any picture of the space-time configurations. We limit ourselves 
to stress that 
images are very similar to the snapshots shown in Section 3 of 
ref.\cite{Rous}, to which we address the reader.}. 
The various dynamical regimes and the phase diagram are reminiscent
of those observed in the 4-neighbours case, although,
as expected, they occur at different values of the parameters.
They are summarized in Table 1~.

\vspace{.5cm}
{\centering \begin{tabular}{|c||c|c|c|c|c|c|c|c|c|c|}
\hline
 &\multicolumn{10}{|l|}{$T_I$}\\
\cline{2-11}
 $T_Q$& 2 & 3 & 4 & 5 & 6 & 7 & 8 & 9 & 10 & 11\\
\hline  
\hline 
1 & P & P & P & D & P & P & D & P & P & P\\
\hline  
2 & P & P & P & D & P & D & D & P & P & F\\
\hline 
3 & P & P & P & P & D & D & D & D & F & F\\
\hline 
4 & P & P & P & E & D & D & D & F & F & F\\
\hline 
5 & P & P & E & E & D & D & F & F & F & F\\
\hline 
6 & P & E & E & E & D & F & F & F & F & F\\
\hline 
7 & E & E & E & E & P & F & F & F & F & F\\
\hline 
8 & E & E & E & E & P & F & F & F & F & F\\
\hline 
9 & E & E & E & E & P & F & F & F & F & F\\
\hline 
10 & E & E & E & E & F & F & F & F & F & F\\
\hline 
\end{tabular}
\par}
\vskip .3cm 
Table 1 - D: disordered evolution, E: extinction, F: ordered front
propagation, P: periodic evolution.
\vskip .5cm 

As shown for the ${\cal N}=4$ case, we have characterized the D-regime by
computing the damage-spreading velocity 
(\ref{voft}), that is found to be a positive
constant for $t$ large enough (e.g, see Figure 1). 

\begin{figure}
\centerline{\psfig{figure=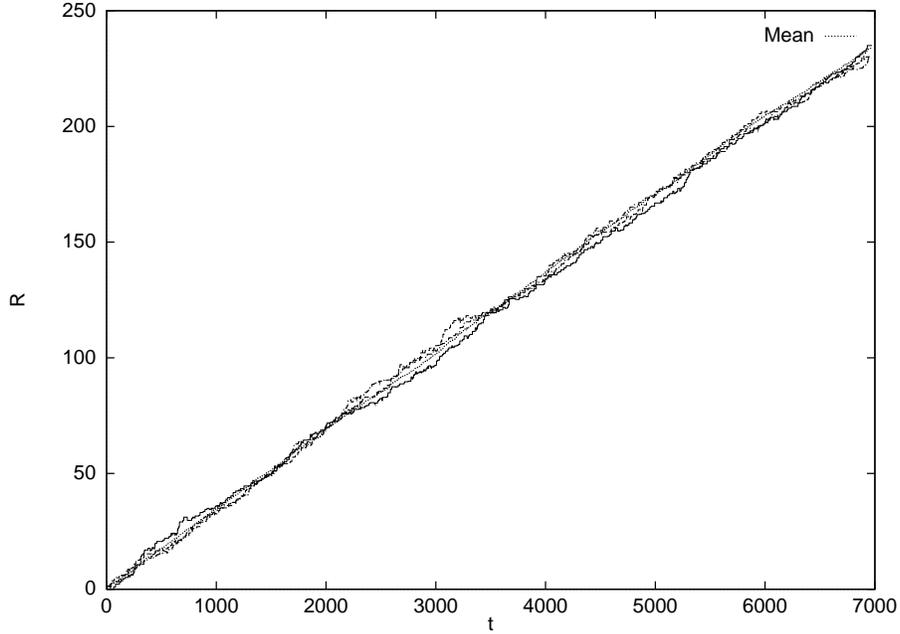,angle=270,width=12cm}}

\caption{$\langle R \rangle$ as a function of $t$ 
for $p=1$ and $T_I=8$, $T_Q=1$ (dotted line), $T_I=7$ ,$T_Q=5$ (full line);
averages are made over 100 initial conditions. 
}
\end{figure}

When starting from a randomly seeded
initial condition, also in the dynamical regimes different from D
the initial randomness may be apparently maintained during the evolution
(apart some cases observed in the P regime, where the space-time 
pattern can be seen to rapidly ``freeze'' into a periodic evolution).
On the other hand, in all of these
cases one finds $\langle v_d \rangle = 0$.  This shows that this quantity is
a reliable order parameter and confirms that the propagation
of information is the main non-linear mechanism yielding unpredictable 
evolution in DCA.

In the following section 
we are going to describe in some detail the effects of variyng $p$ in 
the different regimes, while taking into account in this perspective
measurements of both  $\langle v_d \rangle $ and $b(t)$. 

\section{The Effects of Quenched Phase-Disorder}
  
\subsection{Chaotic Evolution for small $T_Q$ }

For small values of the immunization period $T_Q$
one expects that phase-disorder may not affect the main 
features of a D-regime. In fact, this guess is confirmed
by numerical simulations. Here we consider 
just the case $T_I = 8$ and $T_Q = 1$ as a typical example.
For what concerns the initial state of the cells
we have considered both kinds of initial
condition described in the previous section, verifying that 
they yield the same asymptotic evolution. In particular, the
time-averaged values of the density of $I$-cells and of $Q$-cells
are found to approach asymptotic values independent of the chosen
initial condition. 

We have measured the damage spreading velocity 
$\langle v_d\rangle$ for different values of $p$.  

\begin{figure}
\centerline{\psfig{figure=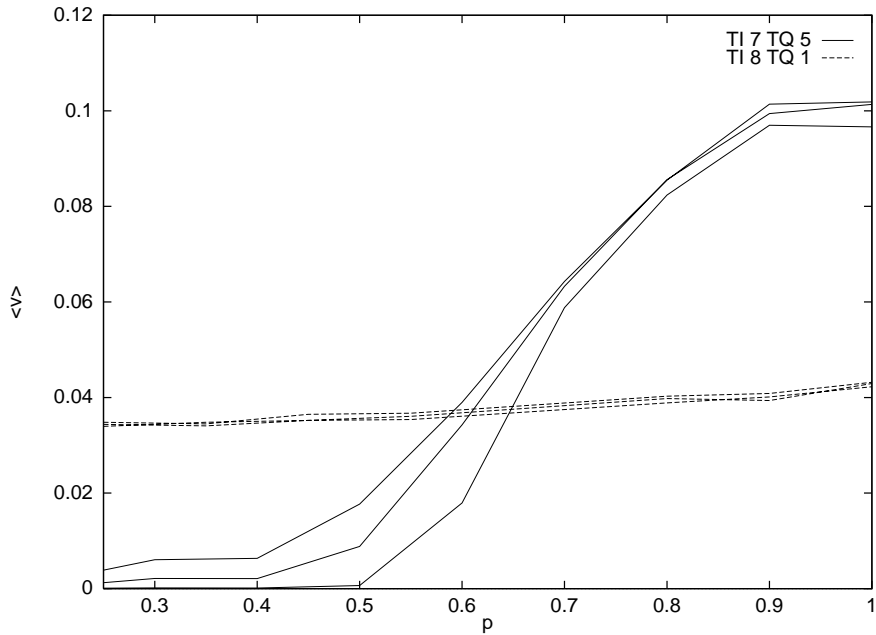,angle=270,width=12cm}} 
\caption{The average damage spreading velocity 
$\langle v_d\rangle$ of the difference pattern 
for $T_I=8$, $T_Q=1$ (dotted lines), $T_I=7$ ,$T_Q=5$ (full lines);
the different lines correspond to ensemble averages performed over
100 initial conditions and time averages performed over the
last 500 points for t=500, 1000 and 2000 (from top to bottom). }
\label{fig:v} 
\end{figure}
 
Fig.2 shows that $\langle v_d \rangle$  has a finite value, that does 
not vary significantly 
in the whole range of $p$. This indicates that phase-disorder, in this 
case, is unable to compete with the mechanism of infection propagation.
Conversely, it is found to progressively remove the anisotropy of the
interaction when $p$ is decreased, thus increasing the efficiency of
the epidemic propagation. This conclusion can be drawn by measuring  
$b(t)$. Specifically, 
the hystogram of the frequency of occurrence of $b$ values vs. $b$ (i.e., the 
normalized probability distribution $P(b)$~) averaged over an hundred of
initial
conditions  is shown in Fig.3.
\begin{figure}
\centerline{\vbox{
\psfig{figure=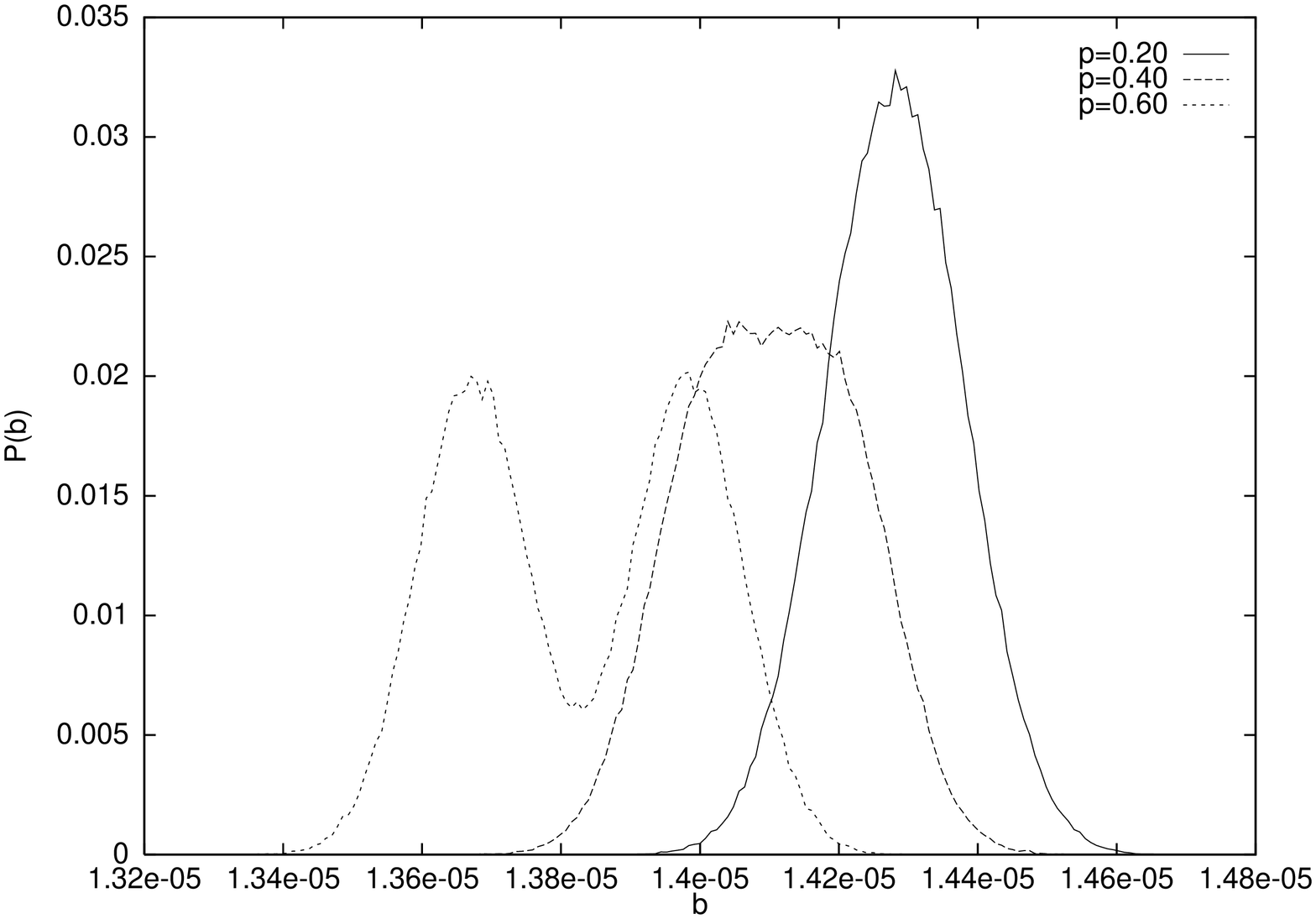,angle=270,width=12cm}
\psfig{figure=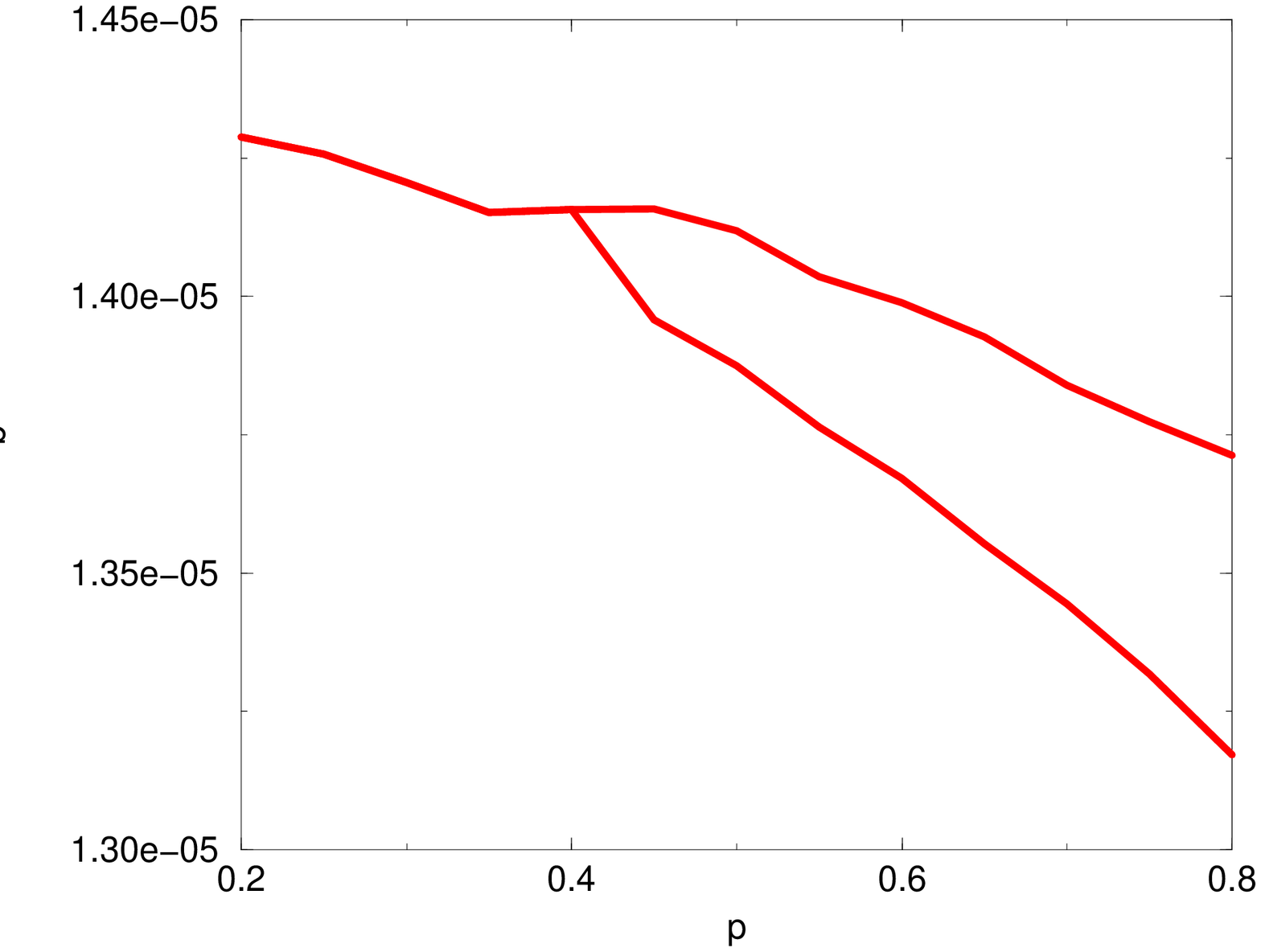,width=12cm}
}}
\caption{At the top: the probability distribution $P(b)$ for $T_I=8$
and $T_Q=1$ and for three different values of $p$; it has been obtained
averaging over 100 initial conditions. At the bottom: the position of the
maxima is reported as a function of $p$, showing that collapse occurs
at $p\approx 0.4$~.}
\label{fig:P(b)} 
\end{figure}

For high values of $p$, including the case $p=1$, $P(b)$ exhibits
two separated peaks. We have
verified that they are in correspondence with infection
propagation along the lattice axis ( the right one ) and 
along the diagonal (the left one ). Decreasing $p$ the
two peaks of the distribution approach one another,
until they merge in a unique distribution at $p \approx 0.4$.

This is definitely the main outcome of our analysis of the D-regime
for small values of $T_Q$:
the restoration of the isostropy of the epidemic
propagation due to phase-disorder occurs at a finite
value of $p$ through an inverse-bifurcation mechanism concerning the
probability distribution $P(b)$. This
is highly non trivial, since the bifurcation applies to a 
"statistical" observable and not just to a single dynamical variable.
Upon these results, one is led to conclude that small immunization
times prevent the possibility of estinghuishing the epidemy,
while disorder can make the mechanism even more efficient by
restoration of isotropy. Actually, when $p$ is lowered 
the unique peak of the distribution displaces towards larger 
values of $b$.
 
\subsection{Chaotic Evolution for large $T_Q$ }

A different scenario, similar to the one associated to the phase
transition described in \cite{Rous} when $p$ varies, is expected to 
be observed in the D-regime for larger values of $T_Q$.
As an example, we discuss here the case $T_I=7 $ and $T_Q = 5$.
The numerical measurements of $\langle v_d\rangle$ versus $p$
(see Fig. 2)
allow us to locate a critical values at $p_c \approx 0.5$~. Below 
$p_c$ the damage spreading velocity $\langle v_d \rangle $ vanishes, 
thus indicating that, 
at variance with the previous case, now there is a dynamical phase,
where disorder is able to localize the infection in some lattice region.
A similar feature is not peculiar of DCA, but it
was found also, for instance, in coupled map lattices, where quenched disorder
is able to inhibit chaotic diffusion \cite{Radons}.

The shape of the curve reported in Fig. 2  gives evidence of a continuous
transition. Note that a numerical determination of the  
critical exponents is practically unfeasible, due to finite
size effects that would demand exceedingly large values of $L$.

On the other hand, in this 
case the hystogram of $P(b)$ exhibits
a two-peak structure that is maintained up to $p_{min} = 0.125$.
Nonetheless, when $p$ is decreased the two peaks tend to approach 
each other, still 
shifting towards higher average values of $b(t)$.
Accordingly, the anisotropy of the epidemic propagation is intrinsic to
this dynamical regime, although the propagation of finite amplitude 
perturbations
for  $p < p_c $ is prevented by localization
effects induced by the phase-disorder.

\subsection{Ordered Evolution for $T_I \approx T_Q$ }
 
Phase-disorder is expected to modify also the dynamical 
features of the ordered regimes. For the sake of space, let us
consider here just the interesting case $T_I=7 $ and $T_Q = 6$, where the
evolution is front-like (F) for $p = 1$. The remarkable result  
is that disorder, now, yields a qualitative change
in the dynamics passing from a front-propagation scenario
to a spatially disordered evolution as soon as $p$ is decreased
below 1.
The average damage spreading velocity $\langle v_d\rangle$ remains null 
for any value of $p$. Nonetheless, phase-disorder is able to prevent  
the extinction of the epidemy, that , due to boundary effects, 
occurs for $p = 1$~. 
In Fig. 4 we show a snapshot of a typical spatial
pattern of this dynamics for $p = 0.2 $~. 

\begin{figure}
\centerline{\hbox{
\psfig{figure=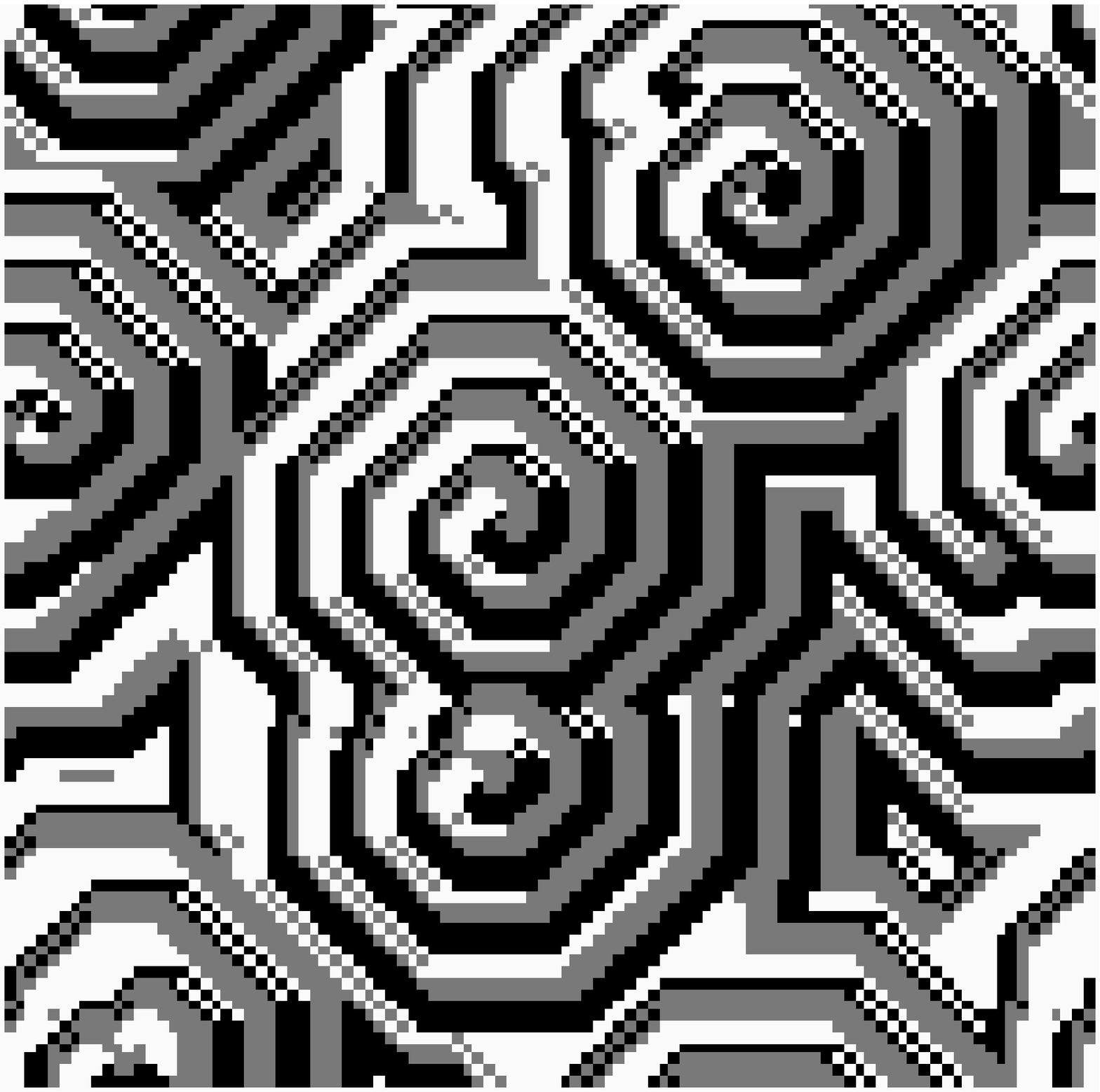,height=5cm,width=5cm}
\hspace{.5cm}
\psfig{figure=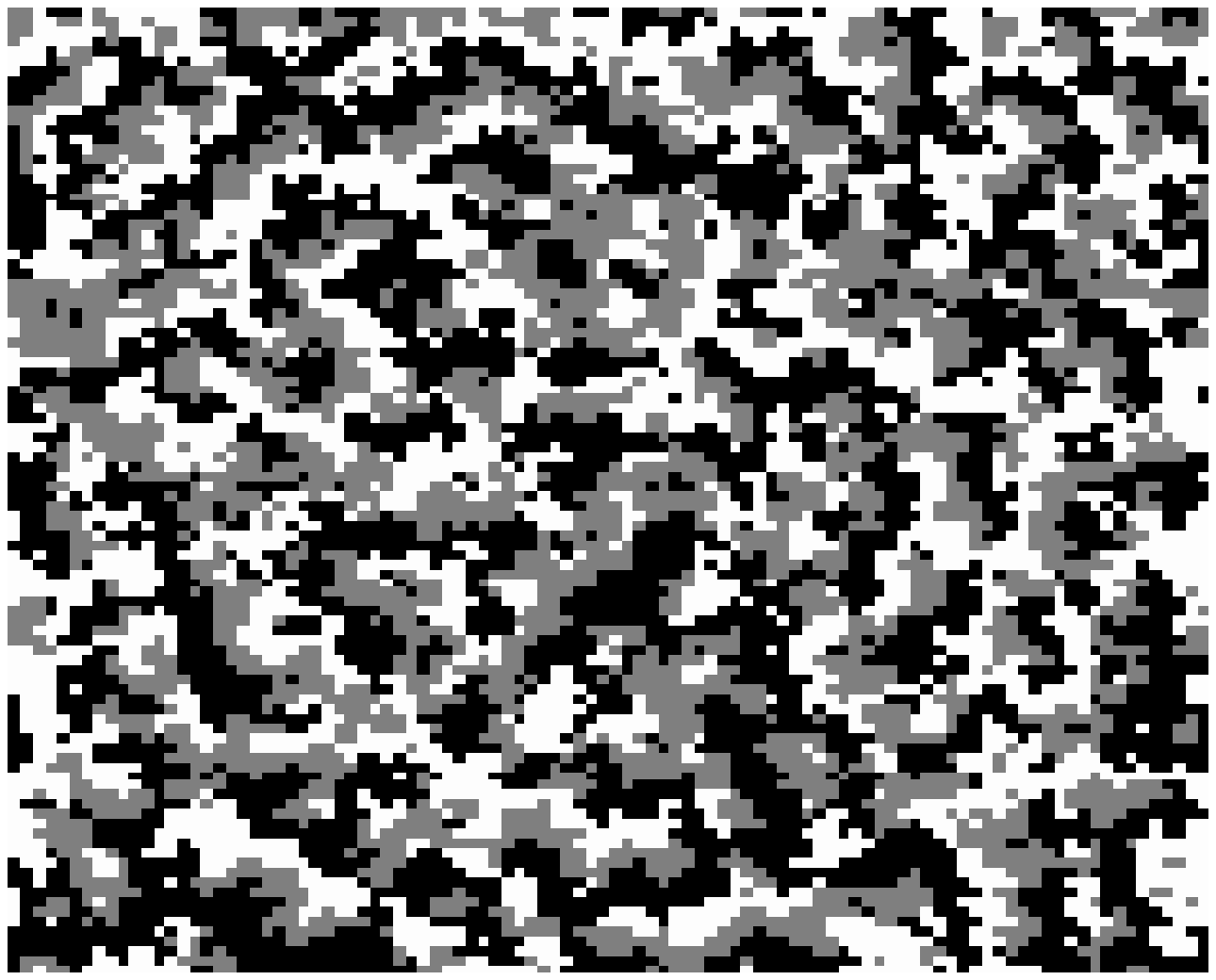,height=5cm,width=5cm}
}}
\vspace{.5cm}
\caption{Snapshot of the pattern for $T_I=7$ and $T_Q=6$,
$p=1.0$ (left) and $p=0.2$ (right) with random initial conditions}
\label{fig:snap1} 
\end{figure}
\begin{figure}
\centerline{
\psfig{figure=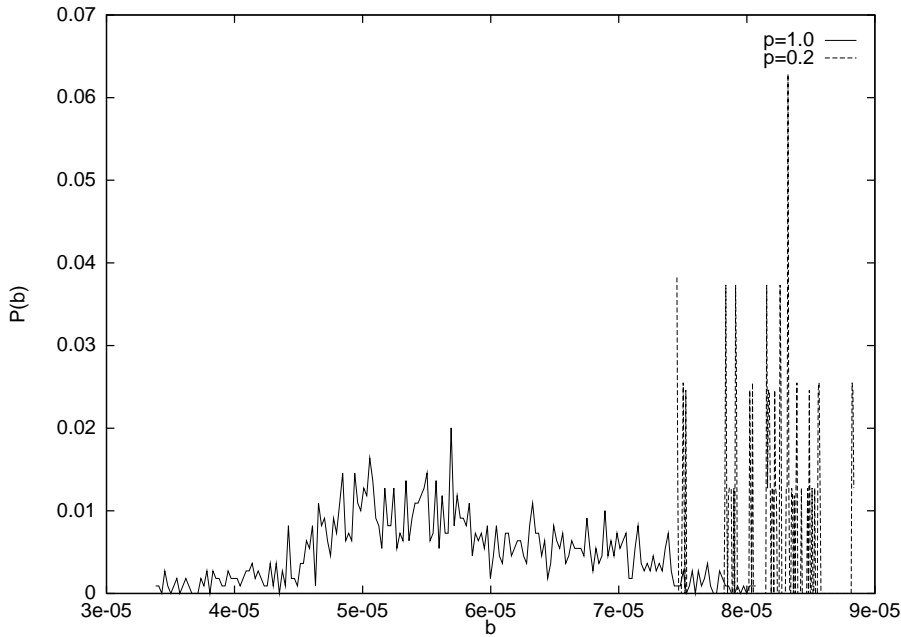,angle=270,width=12cm}
}
\vspace{.5cm}
\caption{Probability 
distributions $P(b)$ for $p=1$ and $p=0.2$. It exhibits a peculiar peaked structure for $p=0.2$.}
\label{fig:peak} 
\end{figure}

When random initial conditions are imposed the system evolves towards 
ordered structures for $p=1$ (Fig. 4, left) and the propagation mechanism 
is described 
by a broad probability distribution $P(b)$ (Fig. 5, full line).
For $p=0.2$ random initial conditions yield disordered evolution quite
similar to one-infected-site initial conditions (Fig. 4, right) 
while $P(b)$ mantains multi-peaked structure (Fig.5, dotted line).
This can be interpreted as an indication of different propagation 
mechanisms characterized by different space-time periodicities.

\section{Conclusions and perspectives}
In this paper we have analyzed the effect of disorder over 
some typical dynamical regimes occurring in an epidemic
propagation 8-neighbour DCA model.

In analogy to what already observed in \cite{Rous}, different dynamical 
regimes including periodic evolution, front
propagation and chaotic patterns are observed when the infection and the 
immunization periods are varied. Here we have performed a more refined
analysis of these regimes, while pointing out new features of suh a class 
of DCA.

In particular, in Sec. 4 we have described three 
chaotic regimes characterized by different dynamical and statistical properties.
Two of them are found to correspond to a D-regime in the phase-synchronous
case. In such cases the presence of quenched phase-disorder may yield very 
different
consequences: either a localization transition for the damage
spreading velocity or an inverse bifurcation in the probability 
distribution of the infection rate.

Conversely, the third case occurs just as an effect of increasing 
phase-disorder applied to the front-propagation regime.
A sort of weakly turbulent phase mediated by defects in the form of spiral 
waves is observed for sufficiently small values of $p$. 
It is worth stressing that this regime is charecterized neither by propagation
of damage nor by a bifurcation-transition in the shape of 
of $P(b)$. In this sense it could not be interpreted as 
a truly chaotic case, indicating that it occurs as a complex regime
located at the border (in the $(T_I , T_Q )$-parameter space) between chaotic
and regular evolution. This scenario is definitely reminiscent of what
has been observed in coupled stable-map models \cite{Cuche,Cecco}.

\acknowledgments
We want to acknowledge useful discussions with B. Giorgini and
G. Rousseau, who contributed to the early stages of this
research. M.B. thanks the Dipartimento di Matematica Applicata 
``G. Sansone'' for friendly hospitality.
We also thank I.S.I. in Torino for the kind hospitality during 
the workshop of the EU HC\&M Network ERB-CHRX-CT940546 on ``Complexity and
Chaos'', where part of this work was performed.
    

\end{document}